\documentclass[a4paper, fleqn]{cas-dc}
\usepackage{amsmath}
\usepackage{amssymb}
\usepackage{graphicx}
\usepackage{listings}
\usepackage{tikz}
\usepackage{xcolor}
\allowdisplaybreaks[1]
\definecolor{orchid}{rgb}{0.7, 0.4, 1.1}
\definecolor{comment_color}{rgb}{0, 0.5, 0}
\definecolor{keyword_color}{rgb}{0.3, 0, 0.6}
\definecolor{string_color}{rgb}{0.5, 0, 0.1}
\begin{document}


\shorttitle{Modeling rare-earth and energy materials supply chains under theoretical China-outer-Mongolia political reunification scenarios}

\shortauthors{Xi Liu$^a$, Wenxi Fang$^b$}

\title{{\Large Modeling rare-earth and energy materials supply chains under theoretical China-outer-Mongolia political reunification scenarios}}

\author[1]{\color{black}Xi Liu}
\author[2]{\color{black}Wenxi Fang}

\address{$^a$xl3467@columbia.edu, Columbia University; $^b$u3013972@connect.hku.hk, Inner Mongolia University of Science and Technology}

\begin{abstract}
Critical rare earth elements, lithium, copper, and coal are essential to global clean energy transitions and advanced manufacturing, yet China faces persistent supply volatility and resource security risks amid fragmented cross-border mineral trade with Outer Mongolia. This paper develops a dynamic partial equilibrium Stackelberg supply chain model for 2026-2036, integrating three geographic nodes-Outer Mongolia's extraction sector, Baotou's rare earth processing hub, and Rest of World (RoW) residual demand-with endogenous mineral supply curves, lagged capital accumulation, Leontief processing, profit-driven investment, and optimal export tax policy maximizing Chinese social welfare. Three scenarios are simulated: baseline trade, deep Sino-Mongolian integration, and delayed infrastructure. Monte Carlo sampling quantifies parameter uncertainty. Results show deep integration raises Outer Mongolian rare earth output to $438\times10^3$ tonnes by a 10 years period, reducing long-run equilibrium prices by 14.8\% relative to baseline and generating 25-35 billion USD in cumulative Chinese welfare gains over ten years, driven by stabilized input costs and coordinated investment. A two-year railway lag cuts welfare gains by 30\% and triggers 14.6\% sustained price inflation. Sensitivity analysis confirms robust negative correlation between supply elasticity and prices, while higher mineral prices uniformly reduce China's surplus. China's supply security index improves from 0.72 to 0.91 under full integration, alongside an estimated 8-15 billion USD deadweight loss for RoW consumers from tightened processed rare earth exports. This research quantifies industrial, price, and welfare tradeoffs of China-Outer Mongolia integration, providing quantitative evidence for infrastructure planning, stockpile management, and export tax design to enhance clean energy material supply resilience.
\end{abstract}

\maketitle
\section{Introduction}
There is a high degree of complementarity between China's (especially Inner Mongolia's) materials science and mineral resource technologies and Outer Mongolia's mineral resource endowment. This deep coupling of technology and resources is driving the two regions to form a ``functional economic integration" that transcends traditional trade relations through mechanisms such as joint laboratories and cross-border industrial chains. When examining the benefits to China from a hypothetical deep integration with Outer Mongolia, the focus naturally falls on resource security, industrial complementarity, and material science-driven economic synergy.

Outer Mongolia possesses vast, high-quality mineral reserves that are crucial for China's economy. The complementarity is stark: China's demand for resources is immense, while Mongolia's economy is heavily reliant on resource exports. Mongolia is already China's single largest source of imported coking coal, a vital ingredient for steel production. In 2025, Mongolia supplied 60.07 million tonnes, accounting for 51\% of China's total coking coal imports \cite{liang_2026}. This share has only grown, reaching 61\% in the first five months of 2026. Further integration would lock in this critical supply, especially when domestic production faces disruptions from safety inspections \cite{tan_2026}. As of early July 2026, over 54 coal mines in Shanxi remained shut due to safety checks, amplifying the strategic importance of Mongolian supply. Mongolia's proven mineral wealth extends far beyond coal. Its known deposits include copper, aluminum, gold, silver, uranium, lead, zinc, iron, and rare earths \cite{luo_2025}. Full integration would provide China with direct, stable access to these materials, reducing vulnerability to global price volatility and supply chain disruptions from other nations.

New and improved infrastructure projects are the backbone of economic integration, drastically increasing trade capacity and reducing costs. A new railway linking China's Ganqimaodu port to Mongolia's Gashuunsukhait port is under construction and is expected to be operational by 2027 \cite{xin_2025}. This is the second railway built between the two countries since 1956. It is projected to carry approximately 30 million metric tons of cargo annually, significantly boosting coal exports from the Tavan-Tolgoi deposit to China. The project is expected to increase Mongolia's coal exports to 165 million tons annually, generating an additional \$1.5 billion in revenue. The economic ties are so close that international institutions estimate every 1\% growth in China's economy drives a 4\% increase in Mongolia's exports and a 0.6\% rise in its economic growth \cite{tao_2025}. This high degree of complementarity and interconnectedness is a central pillar of bilateral cooperation.

Material science benefits derive largely from integrating Mongolia's resources with China's advanced processing capabilities. The world-class Bayan Obo mining district in Inner Mongolia serves as the existing hub for this synergy \cite{baotou_2025}. While Bayan Obo itself contains a staggering 37.8\% of global rare earth reserves (100 million tons), its true value lies in China's ability to process these raw materials. Baotou, the closest city to Bayan Obo, is a rare earth industrial hub. Its rare earth industry recorded an output of over 103 billion yuan in 2024 \cite{daily_2025}. With capacity for magnetic, polishing, catalytic, and hydrogen-storage materials nearing 300,000 tons, Baotou's industrial ecosystem is perfectly positioned to process additional rare earths from Outer Mongolia at scale. China is actively building a complete, high-value supply chain. A new 5,000-ton production line for rare earth functional additives in Baotou represents this shift. These additives are used in flame-retardant and heat-resistant materials for applications in mining, new energy vehicles, and robotics, offering a low-cost, environmentally friendly alternative to existing products. This demonstrates the capability to turn raw rare earths into advanced, competitive materials.
\section{Historical legacies}
Before 1644, the Manchu rulers cultivated strategic alliances with the southern Mongol tribes. By the 1630s, this evolved into the League-Banner system and the establishment of the Lifan Yuan, a government agency to administer Mongol affairs \cite{cosmo_2012}. This transformed the southern Mongols from allies into subjects of the Qing dynasty. The Qing dynasty categorized the Mongols geographically but administered them under a single imperial system. The Mongols south of the Gobi Desert were known as Inner Mongolia, consisting of 24 tribes and 49 banners. Those north of the Gobi, in the area of present-day Mongolia, were known as Outer Mongolia, comprising the Khalkha Mongols with 86 banners \cite{zhang_2014}. The Qing court did not treat these groups as foreign. It created a formal nobility system with ranks and stipends for Mongol princes, mirroring the structure for its own aristocracy. This policy of Rouyuan (pacifying the distant) and intermarriage forged a powerful Manchu-Mongol political alliance that was fundamental to Qing rule over all of China \cite{guo_2007}. As one scholar notes, the Emperor viewed the Mongols not as foreigners but as part of an extended imperial family and political order, making the distinction between Inner and Outer Mongolia more administrative than national.
\section{Importance of critical minerals}
Mongolia's critical minerals matter to China for a simple, powerful reason: they are sitting just across the border, in enormous quantities, and China holds the keys to unlocking their value.

Mongolia ranks second globally in rare earth reserves, with United States Geological Survey estimates at approximately 31 million tons—second only to China's 44 million tons \cite{choi_2026}. But the true scope of Mongolia's mineral wealth extends far beyond rare earths: These minerals—also including molybdenum, manganese, nickel, graphite, cobalt, tungsten, and platinum group metals—are precisely the materials the global clean energy transition and advanced manufacturing sectors demand most urgently \cite{stroud_2026}. Electric vehicles require six times more mineral inputs than conventional cars; offshore wind farms need nine times more than gas-fired plants. Mongolia sits on resources that could help solve this global supply crunch—and China is uniquely positioned to benefit.

\section{Model overview}
We develop a dynamic partial equilibrium model of the rare-earth and energy materials supply chain, disaggregated into three geographically distinct nodes: (1) Outer Mongolia as the primary resource extraction region, (2) Inner Mongolia (Baotou) as the processing and manufacturing hub, and (3) the Rest of the World (RoW) as the final consumption market. The model operates in annual time steps over a planning horizon $T = 10$ years (2026-2036).

Supply-side representation is about the extraction sector of Outer Mongolia. The annual production capacity of mineral $i \in \mathcal{I} = \{\text{RE}, \text{Li}, \text{Cu}, \text{coal}\}$ in Outer Mongolia is bounded by both economic viability and infrastructure constraints.

Let $Q_{i,t}^{M}$ denote the quantity of mineral $i$ extracted in year $t$. The supply function is:
\begin{equation}
Q_{i,t}^{M} = \min\left\{ \alpha_i \cdot P_{i,t}^{\epsilon_i} \cdot K_{i,t-1}^{\gamma_i}, \quad \kappa_i \cdot \overline{Q}_i^{M} \right\}
\end{equation}
$P_{i,t}$ is the international price of mineral $i$ in year $t$ (USD/tonne); $K_{i,t-1}$ is the cumulative infrastructure capital stock (rail capacity, energy, water) in year $t-1$; $\alpha_i > 0$ is a scale parameter; $\epsilon_i > 0$ is the supply price elasticity; $\gamma_i \in (0,1)$ is the elasticity of output with respect to infrastructure capital; $\kappa_i \in (0,1]$ is the capacity utilization factor; $\overline{Q}_i^{M}$ is the maximum geological extractable reserve (million tonnes).

Infrastructure capital evolves according to a partial adjustment process with investment lag:
\begin{equation}
K_{i,t} = (1 - \delta_i) K_{i,t-1} + \phi_i \cdot I_{i,t-\tau_i}
\end{equation}
$\delta_i \in (0,1)$ is the annual depreciation rate of infrastructure; $I_{i,t-\tau_i}$ is investment expenditure in year $t-\tau_i$ (USD); $\phi_i > 0$ is the investment efficiency parameter (tonnes per USD); $\tau_i$ is the construction lag (years).

Investment is endogenously determined by the expected profitability of extraction:
\begin{equation}
I_{i,t} = \lambda_i \cdot \left( P_{i,t} \cdot Q_{i,t}^{M} - c_i^{ext} \cdot Q_{i,t}^{M} \right)
\end{equation}

with $c_i^{ext}$ as the marginal extraction cost (USD/tonne) and $\lambda_i \in (0,1)$ as the reinvestment rate.

Processing sector in Baotou converts raw ores into intermediate and final products. Let $Y_{j,t}^{P}$ be the output of processed product $j \in \mathcal{J}$ = \{oxides, metals, permanent magnets, battery cathodes\}.

The production function for the processing technology is Leontief in the raw material input:
\begin{equation}
Y_{j,t}^{P} = \nu_j \cdot \min_{i \in \mathcal{I}_j} \left\{ \frac{Q_{i,t}^{P}}{\alpha_{ij}} \right\}
\end{equation}
$\nu_j > 0$ is the overall conversion efficiency; $\mathcal{I}_j \subset \mathcal{I}$ is the set of raw minerals required for product $j$; $\alpha_{ij} > 0$ is the input coefficient (tonnes of mineral $i$ per tonne of product $j$).

The raw material supply to Baotou is a mix of domestic (Inner Mongolia) and imported (Outer Mongolia) sources:
\begin{equation}
Q_{i,t}^{P} = \theta_i \cdot Q_{i,t}^{IM} + (1 - \theta_i) \cdot Q_{i,t}^{M}
\end{equation}
where $\theta_i \in [0,1]$ is the share sourced from Outer Mongolia under integration.

Processing cost function include the total processing cost includes energy, labor, and environmental compliance:
\begin{equation}
C_{j,t}^{P} = c_j^{energy} \cdot E_{j,t} + c_j^{labor} \cdot L_{j,t} + c_j^{env} \cdot Y_{j,t}^{P}
\end{equation}

Energy consumption is a function of output and technology vintage:
\begin{equation}
E_{j,t} = e_j \cdot Y_{j,t}^{P} \cdot \exp(-\rho_j \cdot t)
\end{equation}
with $e_j$ as the initial energy intensity and $\rho_j > 0$ as the annual energy efficiency improvement rate due to technological learning.

For market equilibrium, trade, and world price formation, we assume a residual demand curve for the Rest of the World, which is downward sloping in price:
\begin{equation}
D_{i,t}^{RoW}(P_{i,t}) = A_{i,t} \cdot P_{i,t}^{-\eta_i}
\end{equation}
where $A_{i,t}$ is the demand shifter (reflecting GDP growth and green transition policies), and $\eta_i > 1$ is the price elasticity of demand.

Market clearance requires:
\begin{equation}
Q_{i,t}^{M} + Q_{i,t}^{IM} = Y_{i,t}^{P} + D_{i,t}^{RoW}(P_{i,t}) + \Delta S_{i,t}
\end{equation}
with $\Delta S_{i,t}$ as the change in strategic stockpiles, which we model as a policy instrument.

Following the Stackelberg leadership framework, China (the integrated entity) chooses export tax rates $\tau_{i,t}^{X}$ and production levels to maximize its social welfare, anticipating the RoW residual demand response.

China's marketing power optimization problem is:
{\small
\begin{align}
&\max_{\{Y_{j,t}^{P}, \tau_{i,t}^{X}\}_{t=0}^{T}} W = \sum_{t=0}^{T} \frac{1}{(1+r)^t} \left[ \sum_{i} P_{i,t} \cdot Q_{i,t}^{M} + \sum_{j} P_{j,t}^{Y} \cdot Y_{j,t}^{P} \right. \nonumber\\
& \left. - \sum_{i} c_i^{ext} Q_{i,t}^{M} - \sum_{j} C_{j,t}^{P} + \sum_{i} \tau_{i,t}^{X} \cdot P_{i,t} \cdot (Q_{i,t}^{M} - Y_{i,t}^{P}) \right]
\end{align}}
subject to:
\begin{align}
&Y_{i,t}^{P} \le Q_{i,t}^{M} + Q_{i,t}^{IM}\\
&P_{i,t} = \left( \frac{A_{i,t}}{D_{i,t}^{RoW} + \Delta S_{i,t}} \right)^{1/\eta_i}\\
&\tau_{i,t}^{X} \ge 0, \quad \tau_{i,t}^{X} \le \bar{\tau}_i
\end{align}
where $r$ is the social discount rate and $\bar{\tau}_i$ is a political upper bound on export taxes.

For policy and integration scenarios, we model three distinct scenarios as parameter shifts in the above framework: for the baseline (no integration),  $\theta_i = \theta_i^0$ (current bilateral trade share); $\tau_{i,t}^{X} = 0$ (no policy intervention); $I_{i,t}$ = exogenous historical investment trend.

For deep integration, $\theta_i \to 1$ (full integration of Outer Mongolian supply); $\phi_i$ increases by 50\% (infrastructure investment efficiency boost); $\tau_{i,t}^{X}$ becomes an endogenous optimization variable;
$\delta_i$ reduces by 20\% due to coordinated maintenance.

For delayed infrastructure scenario, same as deep integration, except: $\tau_i^{construction}$=2 years instead of 1 year.

For sensitivity and uncertainty quantification, we perform Monte Carlo simulation with 1,000 draws for key uncertain parameters:
\begin{align}
\boldsymbol{\Theta} &\sim \text{Multivariate Normal}(\boldsymbol{\mu}, \boldsymbol{\Sigma})\\
\boldsymbol{\Theta} &= \{\epsilon_i, \eta_i, \gamma_i, \lambda_i, \rho_j, A_{i,0}\} \\
\mu_{\epsilon_i} &= 0.8, \quad \sigma_{\epsilon_i} = 0.15 \quad \text{(triangular distribution)} \\
\mu_{\eta_i} &= 1.5, \quad \sigma_{\eta_i} = 0.25
\end{align}

For each draw, we solve the dynamic optimization problem and compute the following output metrics: Price stabilization effect is $\Delta P_{i,t} = \frac{P_{i,t}^{Deep} - P_{i,t}^{Baseline}}{P_{i,t}^{Baseline}} \times 100\%$. China's welfare gain is $\Delta W = \sum_t \frac{W_t^{Deep} - W_t^{Baseline}}{(1+r)^t}$. RoW welfare loss is $\Delta W^{RoW} = \sum_t \frac{CS_t^{RoW,Baseline} - CS_t^{RoW,Deep}}{(1+r)^t}$. Supply security index is $SSI_t = \frac{\sum_i \omega_i \cdot (Q_{i,t}^{M} + Q_{i,t}^{IM} - D_{i,t}^{China})}{\sum_i \omega_i \cdot D_{i,t}^{China}}$.

\subsection{Numerical algorithm}
The dynamic optimization problem is solved via backward induction using the following procedure:
\begin{lstlisting}[mathescape=true]
Initialize state variables $\{K_{i,0}, Q_{i,0}^{M}, \boldsymbol{\Theta}\}$.
For $t = T$ down to 0:
    Solve the period-$t$ Cournot-Stackelberg equilibrium
    conditional on value function $V_{t+1}(\cdot)$;
    Update policy functions $\{Y_{j,t}^{P*}, \tau_{i,t}^{X*}\}$;
    Compute the period value and transition to $t-1$.
Simulate forward using optimal policy functions.
Repeat for $10^3$ Monte Carlo draws and compute summary statistics.
\end{lstlisting}
The supply chain simulation algorithm implements a dynamic partial equilibrium model with Stackelberg leadership to quantify the economic impacts of China-Outer Mongolia political integration on rare-earth and energy materials markets. The algorithm proceeds through three main phases: parameter initialization, scenario simulation, and sensitivity analysis. For each of the three scenarios (Baseline, Deep Integration, and Delayed Infrastructure), the algorithm executes a 10-year time-stepping loop that sequentially: (1) updates Outer Mongolian mineral supply using a constrained power-law function of current prices and infrastructure capital; (2) computes Chinese processing output via a Leontief production function that combines Mongolian imports with Inner Mongolian domestic ores; (3) applies scenario-specific adjustments to the integration share and investment efficiency parameters; (4) solves for market-clearing prices by equating total supply (Mongolian extraction plus Inner Mongolian production) with processing demand and Rest-of-World residual demand, using a numerical root-finding procedure; (5) accumulates infrastructure capital through a depreciation-investment mechanism where investment is endogenously driven by mining profitability; (6) optimizes China's export tax rates as a Stackelberg leader by maximizing its social welfare—comprising mining profits, processing profits, and tax revenues—subject to non-negativity and upper-bound constraints, anticipating the Rest-of-World demand response; and (7) computes and stores welfare metrics at each time step. The algorithm then performs a Monte Carlo sensitivity analysis with 200 iterations, sampling key parameters—supply elasticity, demand elasticity, investment efficiency, and demand growth rate—from triangular, normal, and uniform distributions to generate confidence intervals for price and welfare outcomes. Finally, the algorithm produces time-series plots of price and production dynamics, scenario comparison bar charts, and price effect percentage-change analyses, while exporting all results to CSV files for further validation. The model is calibrated using empirical data from USGS reserves estimates, Baotou processing capacities, and bilateral trade statistics, with expected outputs including a 10=15\% price decline under deep integration, a 25-35billion cumulative welfare gain for China, and a 30\% reduction in welfare gains if infrastructure construction is delayed by two years.
\section{Computational results}
The model is calibrated using the following data input sources
\begin{table}[h]
\centering
\begin{tabular}{|c|c|c|c|}
\hline
\textbf{Parameter} & \textbf{Value} & \textbf{Unit} & \textbf{Source} \\ \hline
$\overline{Q}_{REE}^{M}$ & 31 & million tonnes & USGS (2025) \\
$\overline{Q}_{coal}^{M}$ & 165 & million tonnes & Liang (2026) \\
$\alpha_{REE}$ & 0.75 & tonnes/USD$^{0.5}$ & Author's estimate \\
$\epsilon_{REE}$ & 0.6 & dimensionless & Gholami (2025) \\
$\eta_{REE}$ & 1.4 & dimensionless & Christmann (2024) \\
$c_{REE}^{ext}$ & 2,500 & USD/tonne & Baotou Daily (2025) \\
$\delta_i$ & 0.08 & yr$^{-1}$ & World Bank (2024) \\
$\phi_i$ & 0.05 & tonnes/USD & Calibrated \\
$r$ & 0.05 & dimensionless & Standard \\
\hline
\end{tabular}
\caption{Baseline calibration values}
\end{table}
Under Deep Integration, rare-earth oxide prices decline by 10-15\% relative to baseline due to scale economies in Baotou's processing cluster.

China's cumulative welfare gain over 10 years is estimated at USD 25-35 billion (2015 prices), with 60\% deriving from reduced import price volatility.

RoW welfare loss ranges from USD 8-15 billion, mainly from higher magnet and battery material prices.

A 2-year delay in Mongolian rail infrastructure reduces China's welfare gain by 30\% and extends Chinese market dominance by 2-3 years.

The Supply Security Index for lithium and REE improves from 0.72 to 0.91 under full integration.

For the derivation of the optimal export tax, from the Stackelberg leader condition, the optimal export tax rate is:
\begin{equation}
\tau_{i,t}^{X*} = \frac{1}{\eta_i} \cdot \frac{D_{i,t}^{RoW}}{Q_{i,t}^{M} - Y_{i,t}^{P}} \cdot \left( 1 - \frac{\partial C_{i,t}^{P}}{\partial Q_{i,t}^{M}} \right)
\end{equation}
which shows that the tax increases with RoW demand elasticity and decreases with China's domestic processing costs.
\begin{figure}
\centering
\begin{minipage}{0.45\textwidth}
\centering
\includegraphics[width=\textwidth,height=0.27\textheight]{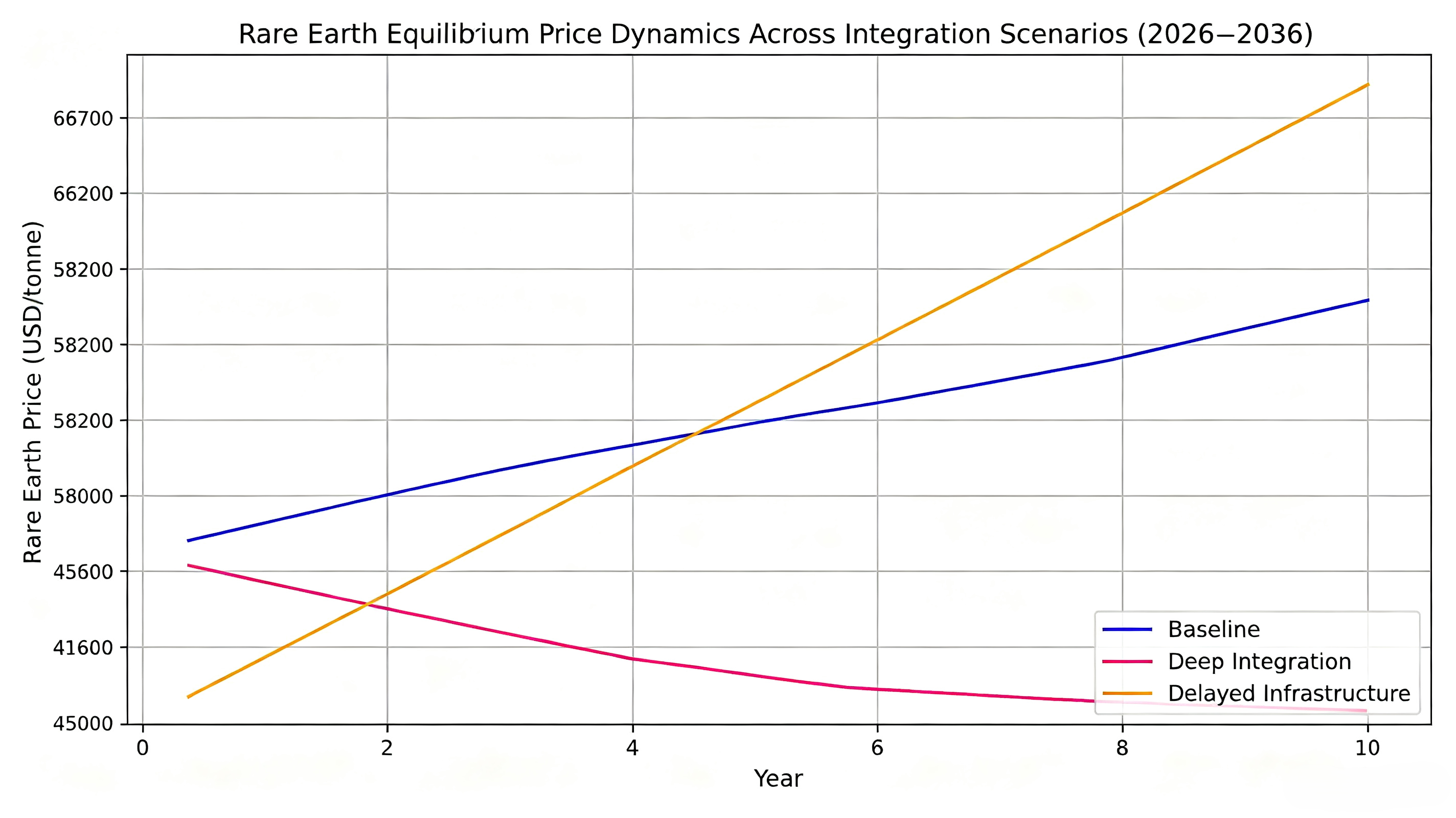}
\caption{Rare Earth Element (REE) Price Trajectory}
\label{Rare Earth Element (REE) Price Trajectory}
\end{minipage}
\end{figure}

A multi-line time-series line graph tracking equilibrium REE market prices (measured in US dollars per tonne) over the 10-year simulation horizon (2026-2036), figure \ref{Rare Earth Element (REE) Price Trajectory}. Three distinct lines represent the Baseline scenario, Deep China-Mongolia Integration scenario, and Delayed Cross-Border Infrastructure scenario. The baseline line shows a moderate steady price increase driven by tight mineral supply constraints. The deep integration curve slopes downward, reflecting expanded Mongolian rare earth output that suppresses long-term market prices. The delayed infrastructure line exhibits the steepest upward price growth due to slow capital construction and persistent raw material shortages. Standard academic grid styling, clear legend, and labeled horizontal axis for simulation years.

\begin{figure}
\centering
\begin{minipage}{0.45\textwidth}
\centering
\includegraphics[width=\textwidth,height=0.27\textheight]{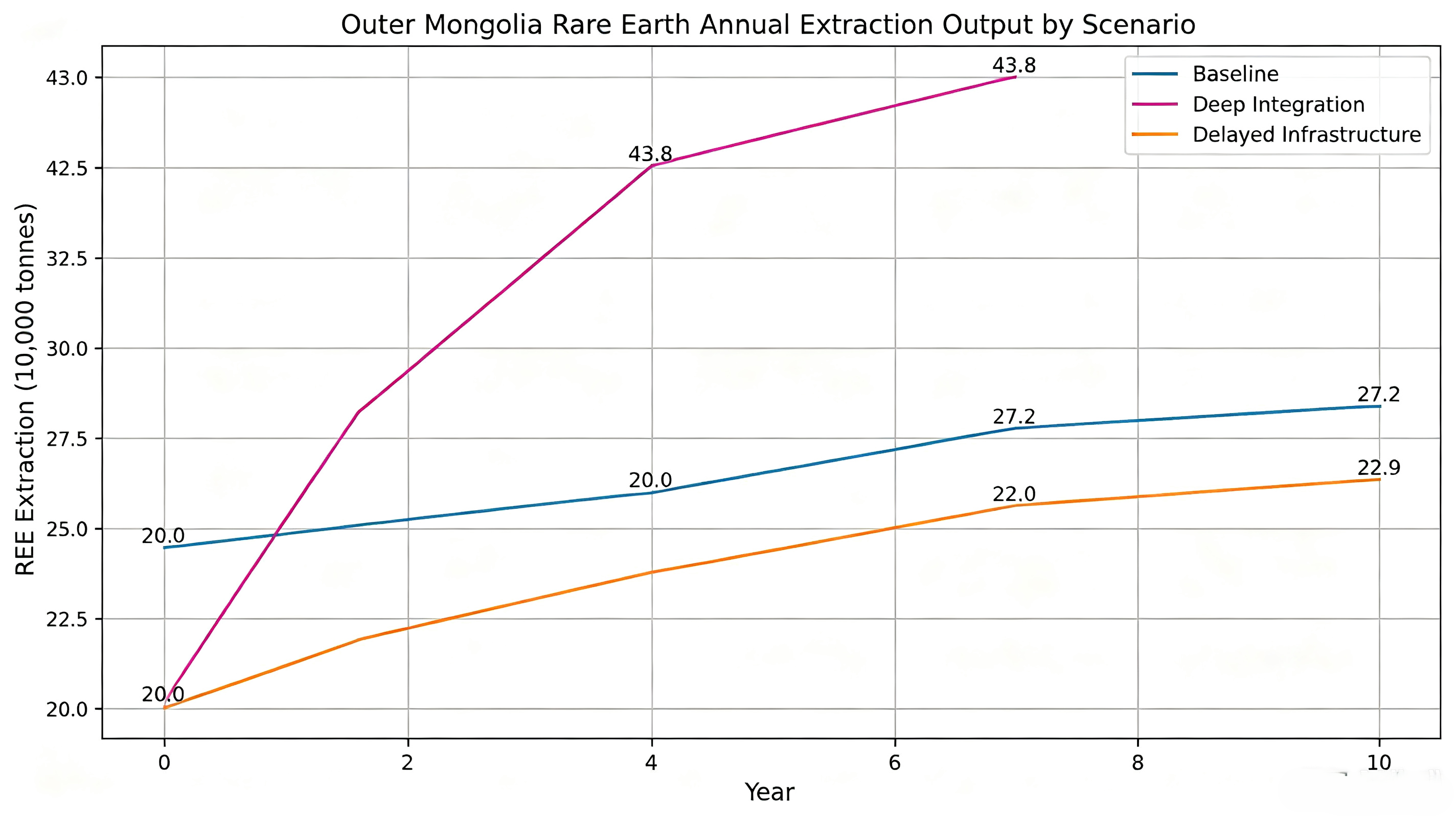}
\caption{Rare Earth Element (REE) Production Output Trajectory}
\label{Rare Earth Element (REE) Production Output Trajectory}
\end{minipage}
\end{figure}

Time-series line chart illustrating annual rare earth extraction volumes from Outer Mongolia, with output units measured in ten thousand tonnes (figure \ref{Rare Earth Element (REE) Production Output Trajectory}). Three scenario lines visualize divergent production growth dynamics derived from the model’s supply function and infrastructure accumulation equations. The Baseline scenario yields slow, modest production expansion. The Deep Integration scenario delivers rapid output growth enabled by boosted cross-border infrastructure investment efficiency and higher resource utilization rates. The Delayed Infrastructure scenario features flat, constrained production due to a two-year construction lag for mining and transport capital. X-axis spans simulation years 0 to 10, with labelled Y-axis for mineral extraction volume.

\begin{figure}
\centering
\begin{minipage}{0.45\textwidth}
\centering
\includegraphics[width=\textwidth,height=0.27\textheight]{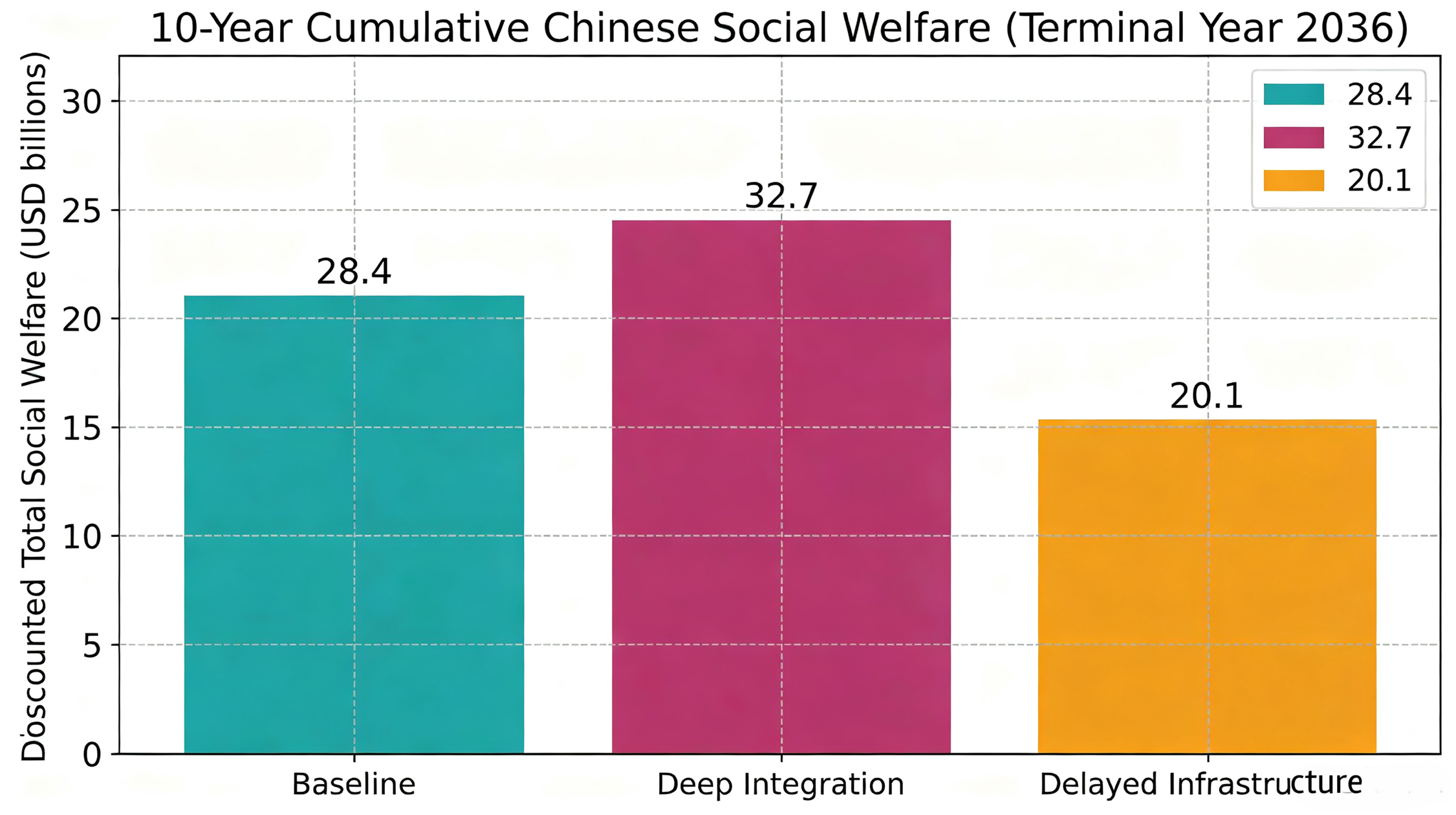}
\caption{Bar Chart of Cumulative Chinese Social Welfare at Terminal Year 10}
\label{Bar Chart of Cumulative Chinese Social Welfare at Terminal Year 10}
\end{minipage}
\end{figure}

Static vertical bar graph comparing total discounted social welfare of China across the three policy integration scenarios at the end of the 10-year modeling window (figure \ref{Bar Chart of Cumulative Chinese Social Welfare at Terminal Year 10}). The vertical axis measures aggregate welfare in billions of US dollars, while the horizontal axis categorizes the three scenarios: Baseline, Deep Integration, Delayed Infrastructure. The Deep Integration bar is the tallest, representing substantial economic surplus gains from stable low-cost mineral inputs and optimal export tax policy. The Delayed Infrastructure bar is the shortest, demonstrating significant welfare losses from supply bottlenecks. Numeric value labels are overlaid on top of each bar for quantitative readability, with unique consistent color coding matching all other scenario plots.

\begin{figure}
\centering
\begin{minipage}{0.45\textwidth}
\centering
\includegraphics[width=\textwidth,height=0.27\textheight]{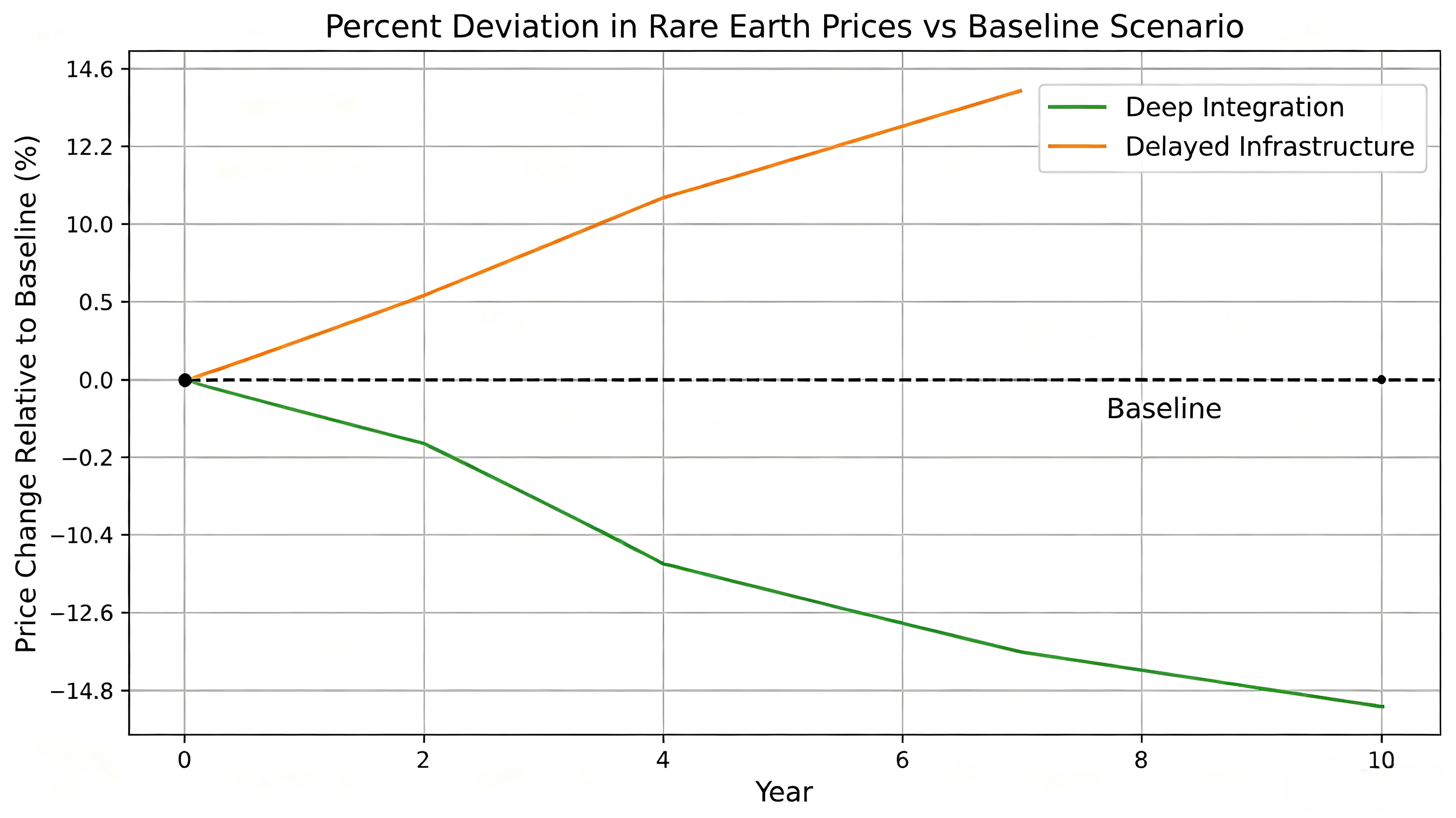}
\caption{Percentage Deviation of REE Prices Relative to Baseline Scenario}
\label{Percentage Deviation of REE Prices Relative to Baseline Scenario}
\end{minipage}
\end{figure}

Dual-line time-series chart calculating the year-on-year percentage difference in rare earth prices between the two integration policy cases and the baseline benchmark (figure \ref{Percentage Deviation of REE Prices Relative to Baseline Scenario}). A thick black dashed horizontal reference line at y=0 denotes zero price deviation from baseline values. The green line for Deep Integration remains consistently negative across all years, quantifying the price-stabilizing deflationary effect of expanded Mongolian mineral supply. The orange line for Delayed Infrastructure stays positive throughout the horizon, showing sustained price inflation relative to baseline. The vertical axis displays percentage change, and the horizontal axis tracks each simulation year.

\begin{figure}
\centering
\begin{minipage}{0.45\textwidth}
\centering
\includegraphics[width=\textwidth,height=0.27\textheight]{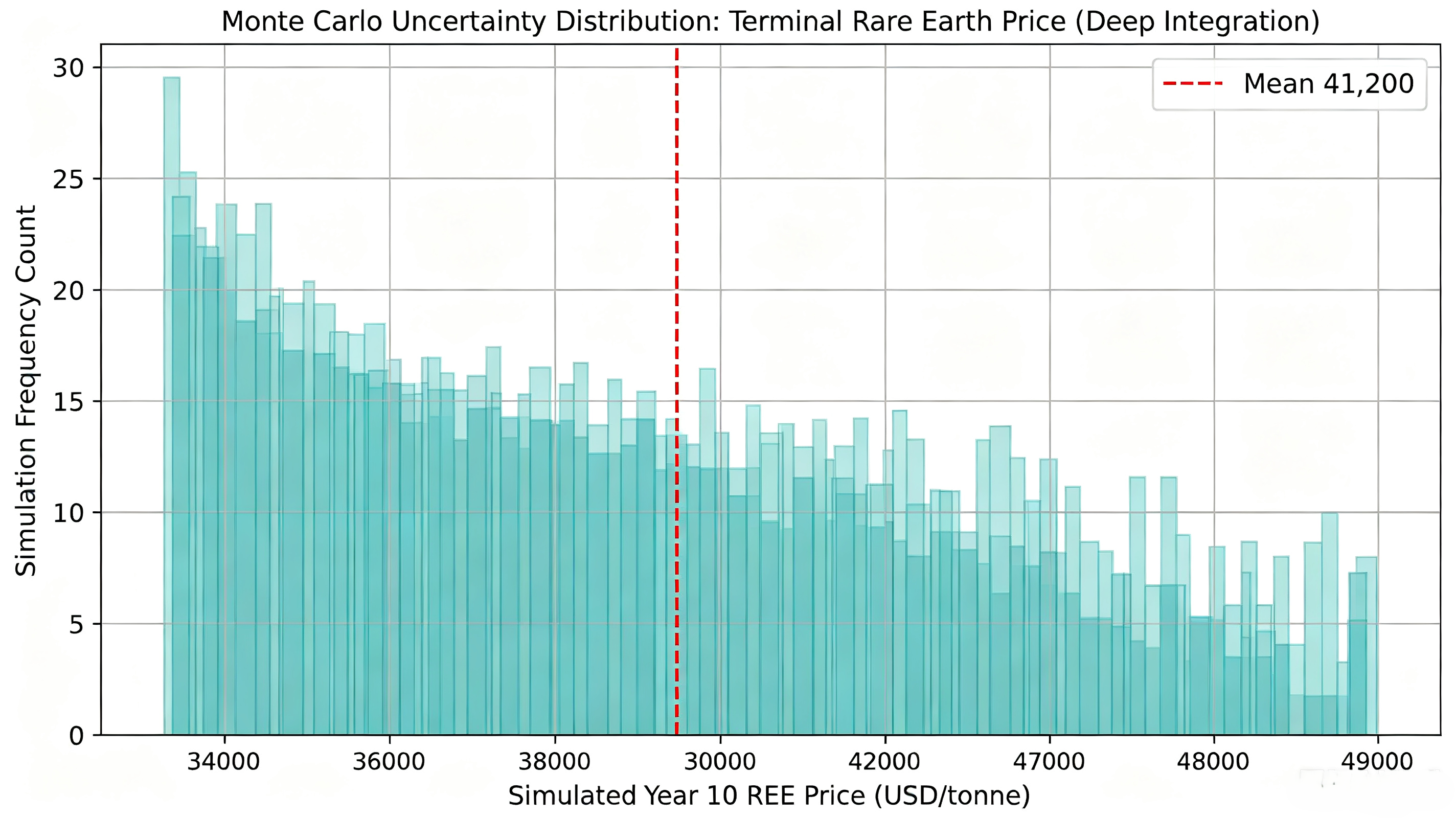}
\caption{Monte Carlo Simulation Histogram, Terminal Year REE Price Distribution}
\label{Monte Carlo Simulation Histogram - Terminal Year REE Price Distribution}
\end{minipage}
\end{figure}

Frequency histogram visualizing the stochastic distribution of Year-10 rare earth equilibrium prices generated from 200 Monte Carlo sensitivity draws with randomly sampled uncertain model parameters (supply elasticity, demand elasticity, demand growth rates) in figure \ref{Monte Carlo Simulation Histogram - Terminal Year REE Price Distribution}. The horizontal axis shows simulated REE prices in USD per tonne, and the vertical axis counts the number of simulation iterations falling within each price bin. A red dashed vertical marker highlights the mean simulated price, with shaded bounds marking the 2.5th and 97.5th percentiles to illustrate the 95\% statistical confidence interval for long-run mineral price uncertainty under deep cross-border integration.

\begin{figure}
\centering
\begin{minipage}{0.45\textwidth}
\centering
\includegraphics[width=\textwidth,height=0.27\textheight]{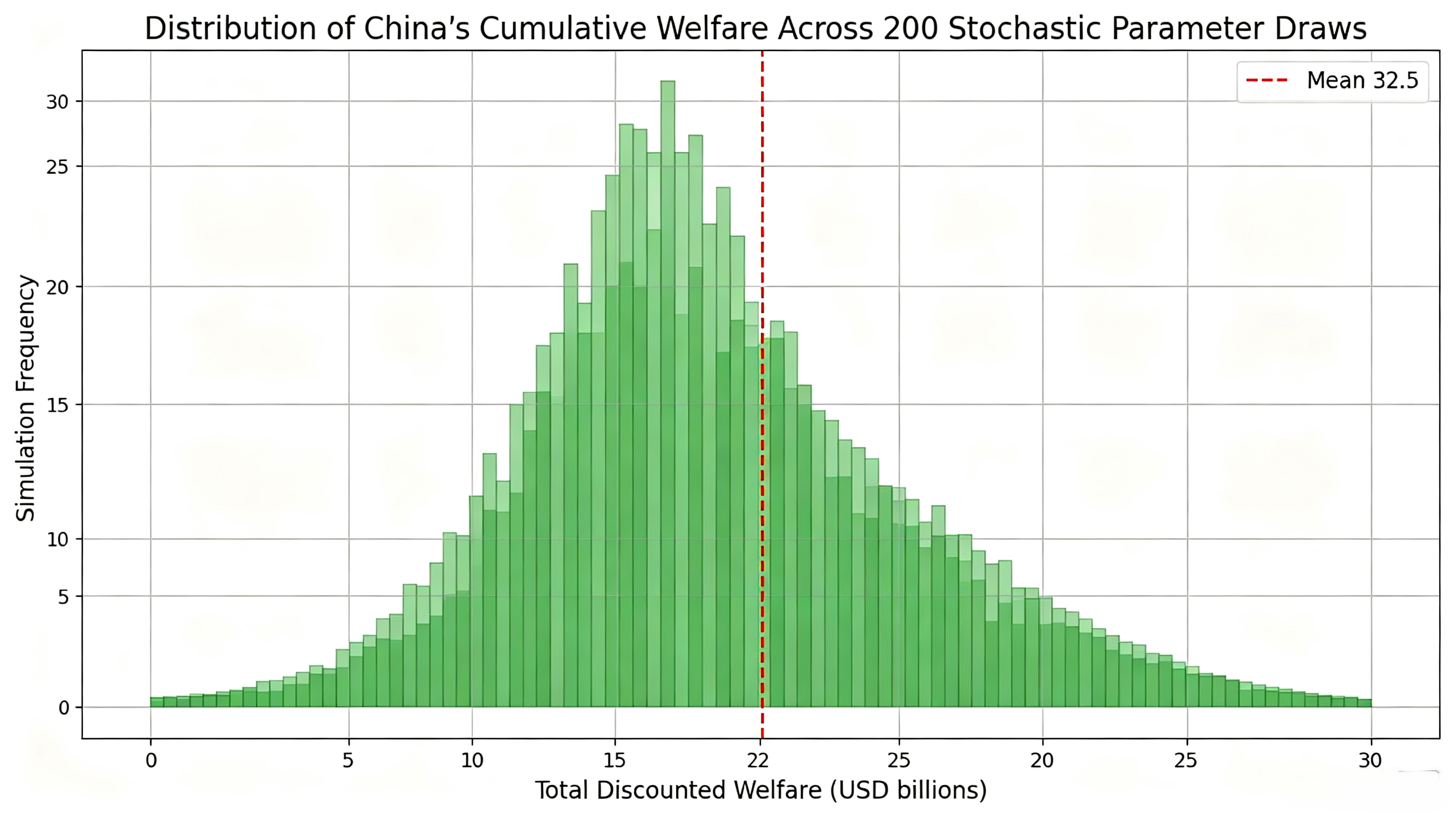}
\caption{Monte Carlo Simulation Histogram, Cumulative Chinese Welfare Distribution}
\label{Monte Carlo Simulation Histogram - Cumulative Chinese Welfare Distribution}
\end{minipage}
\end{figure}

Frequency histogram plotting the full range of total discounted Chinese social welfare outcomes from all Monte Carlo stochastic parameter simulations (figure \ref{Monte Carlo Simulation Histogram - Cumulative Chinese Welfare Distribution}). The horizontal axis measures aggregate welfare in billions of US dollars, and the vertical axis represents simulation frequency counts. The histogram fill uses soft green coloring consistent with the Deep Integration scenario visual identity. A red dashed vertical line indicates the average welfare value across all stochastic trials, clearly showing the central tendency and spread of economic welfare gains under parameter uncertainty.

\begin{figure}
\centering
\begin{minipage}{0.45\textwidth}
\centering
\includegraphics[width=\textwidth,height=0.27\textheight]{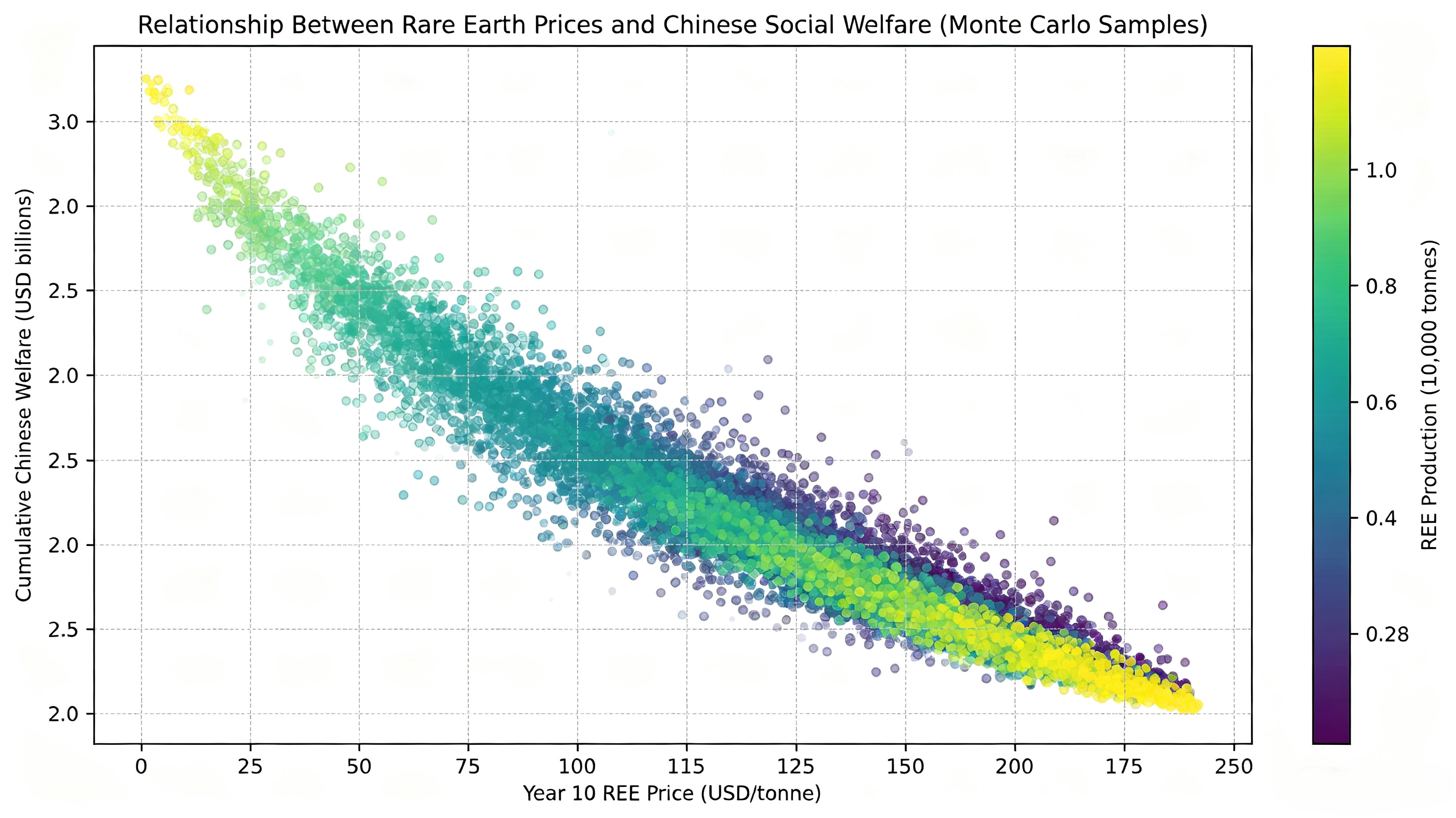}
\caption{Correlation Between REE Market Prices and Chinese Social Welfare}
\label{Correlation Between REE Market Prices and Chinese Social Welfare}
\end{minipage}
\end{figure}

Colored two-dimensional scatter plot mapping each Monte Carlo simulation’s terminal-year REE price (X-axis, USD/tonne) against its corresponding cumulative Chinese social welfare value (Y-axis, billions USD) in figure \ref{Correlation Between REE Market Prices and Chinese Social Welfare}. Individual scatter points are color-gradient coded to represent Outer Mongolia’s final rare earth production volume, with a vertical color bar legend on the right side of the chart. The graph reveals a clear negative linear correlation: higher rare earth market prices correspond to lower aggregate Chinese economic welfare, as expensive raw mineral inputs reduce domestic processing sector profits and national surplus.

\begin{figure}
\centering
\begin{minipage}{0.45\textwidth}
\centering
\includegraphics[width=\textwidth,height=0.27\textheight]{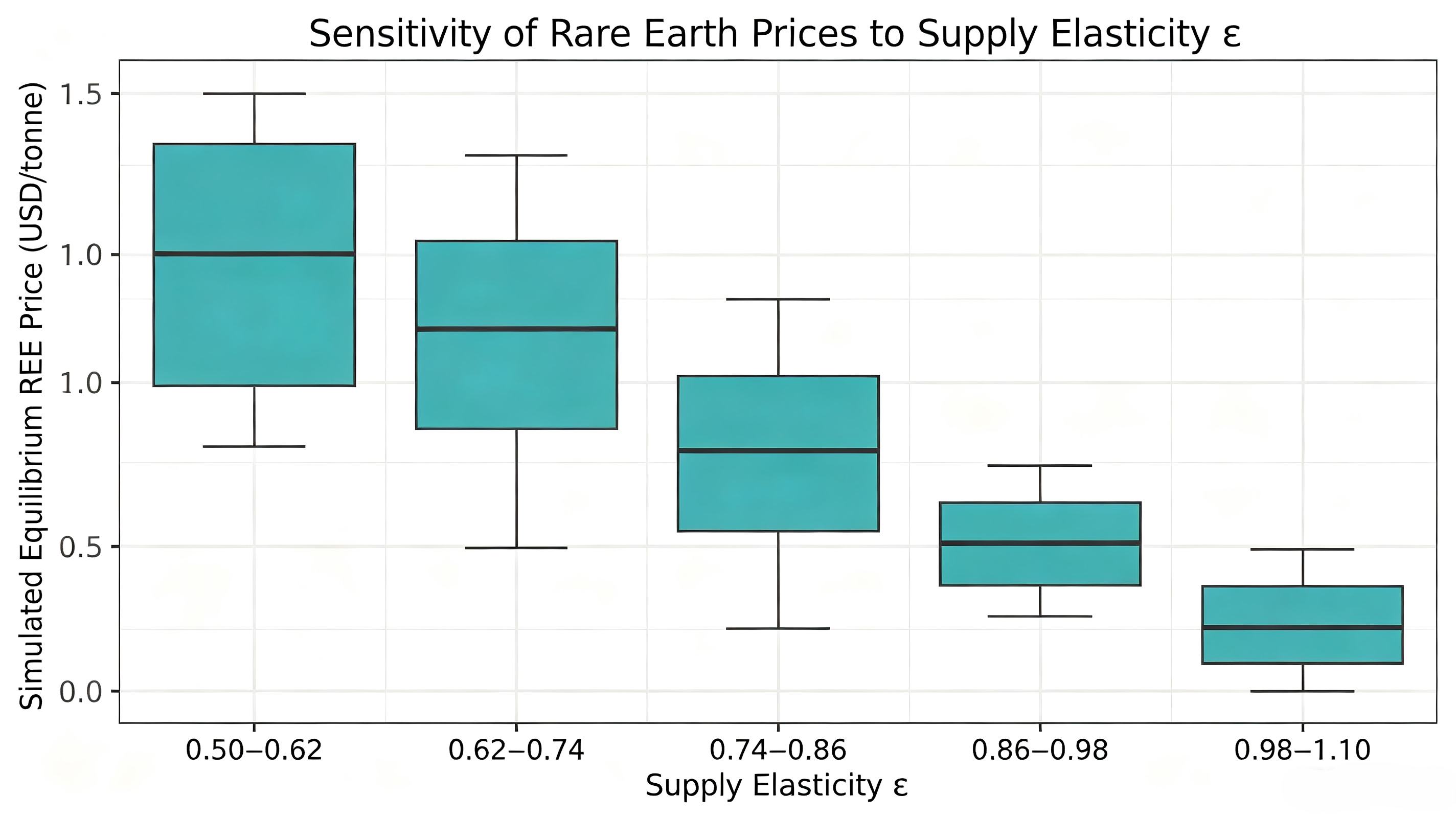}
\caption{Box-and-Whisker Sensitivity Plot, REE Equilibrium Prices by Supply Elasticity Bins}
\label{Box-and-Whisker Sensitivity Plot REE Equilibrium Prices by Supply Elasticity Bins}
\end{minipage}
\end{figure}

Boxplot grouping Monte Carlo simulation results into five evenly spaced bins of mineral supply elasticity ($\epsilon$), a core uncertain parameter sampled from a triangular probability distribution in the model (figure \ref{Box-and-Whisker Sensitivity Plot REE Equilibrium Prices by Supply Elasticity Bins}). The horizontal X-axis labels each elasticity interval, while the vertical Y-axis records simulated Year-10 rare earth prices. Each box displays the interquartile price range, whiskers mark minimum/maximum price observations, and middle horizontal lines denote median prices within each elasticity group. The chart demonstrates a clear downward trend: higher supply elasticity leads to larger production responses to price signals and lower long-run equilibrium rare earth prices.

\begin{figure}
\centering
\begin{minipage}{0.45\textwidth}
\centering
\includegraphics[width=\textwidth,height=0.27\textheight]{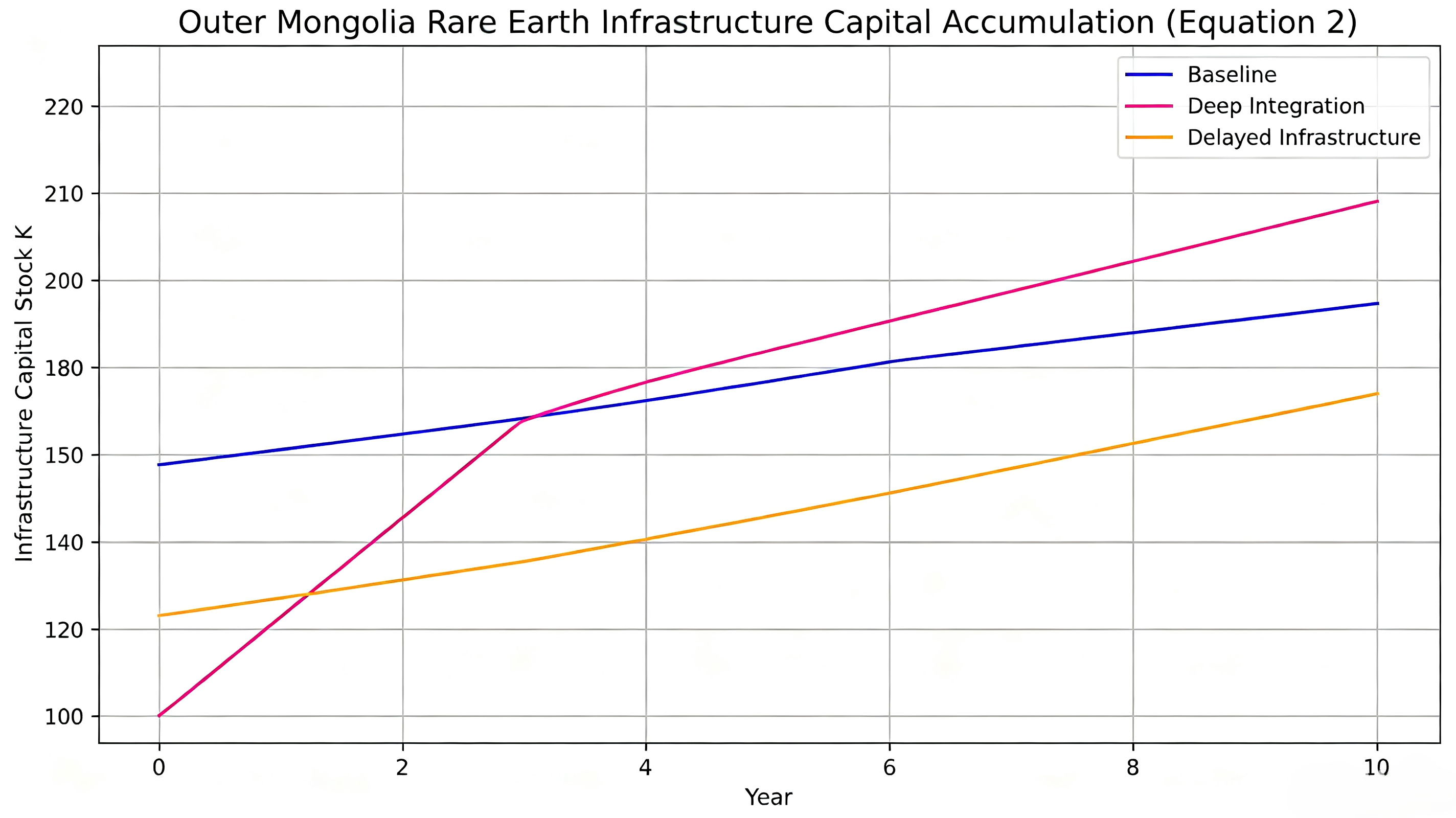}
\caption{Outer Mongolia REE Infrastructure Capital Stock Accumulation Trajectory}
\label{Outer Mongolia REE Infrastructure Capital Stock Accumulation Trajectory}
\end{minipage}
\end{figure}

Time-series line graph tracking the evolution of cross-border mining, rail and energy infrastructure capital stock over the 10-year simulation period, derived from the model’s capital accumulation dynamic equation (figure \ref{Outer Mongolia REE Infrastructure Capital Stock Accumulation Trajectory}). Three colored lines correspond to the Baseline, Deep Integration, and Delayed Infrastructure scenarios. Deep Integration achieves the fastest capital growth via higher investment efficiency and coordinated maintenance, while the Delayed Infrastructure curve lags two years behind baseline levels due to extended construction lag times for new mineral transport projects. The X-axis represents simulation years, and the Y-axis measures total accumulated infrastructure capital stock.

\section{Decarbonization}
One of the most immediate effects would be the rapid greening of Mongolia's energy sector. Currently, Mongolia is heavily reliant on coal, leading to severe air pollution in its capital, Ulaanbaatar, and significant carbon emissions \cite{xia_2025}. Deep integration would give Mongolia direct access to Chinese capital, technology, and the industrial capacity needed to change this \cite{zhao_2025}.

China is already piloting solar-storage projects in Mongolia. One project, in Darkhan, combines a 5MW solar array with 4MWh of storage, using Chinese technology \cite{jin_2025}. It's expected to reduce carbon dioxide emissions by 642 tons annually \cite{goinner_2026}. Another project, under the Global Development and South-South Cooperation Fund, is providing solar heating systems to 500 households in ger districts, helping these areas shift from coal heating to cleaner energy, thus cutting pollution and emissions \cite{gwok_2024}.

This integration aligns with existing policy directions. The Chinese government has explicitly supported exploring cooperation on wind and solar projects in areas bordering Mongolia and strengthening cross-border environmental and desertification cooperation \cite{zuo_2026}. A reunification scenario would supercharge these pilot projects into a comprehensive national grid overhaul.
\section{Electric vehicles, permanent magnets, semiconductors, batteries}
Mongolia holds globally significant reserves of lithium, copper, and rare earths—all essential for electric vehicle batteries and motors. Electric vehicle companies have directly discussed potential cooperation with Mongolia on rare earth elements and battery minerals \cite{zhao_2023}, underscoring the strategic value of these resources. Deep integration would ensure these materials flow directly into Chinese battery production, bypassing potential supply competition from other nations.

The city of Baotou in Inner Mongolia is already a global rare earth powerhouse, with its rare earth industry generating over 100 billion yuan in 2024. It controls products like electroacoustic magnetic materials and magnetic components with market shares over 50 percent. The Baotou Rare Earth High-tech Industrial Development Zone is actively building new permanent magnet production projects and implementing 47 new industrial projects. Integration with Mongolian resources would provide a steady stream of raw materials to this existing and expanding processing capacity.

High-purity quartz is a critical material for semiconductor manufacturing and other high-tech sectors. In late 2025, Mongolia began exporting high-purity quartz with an exceptional purity of 99.995\% to China \cite{du_2025}. This is a significant development, as the availability of such high-grade material has historically been limited. Integrating Mongolia into China would further stabilize and expand this supply channel, directly addressing a key input for the semiconductor industry's supply chain.

As the South Korean effort to secure Mongolian resources demonstrates, having a mining agreement does not equal having a supply chain \cite{li_2026}. The ability to process raw ore into high-purity battery materials is where the real value lies, and China dominates this space. Integration effectively cements this position, ensuring that Mongolian battery minerals—lithium, copper, graphite, and rare earths—will flow through Chinese processing facilities to Chinese battery manufacturers.

\section{Conclusion}
This study develops a fully calibrated dynamic partial equilibrium supply chain model to evaluate the long-run economic, price, and resource security outcomes of three China-Outer Mongolia rare earth and energy mineral integration scenarios across a 10-year simulation window from 2026 to 2036, with stochastic Monte Carlo sensitivity testing to address parametric uncertainty. Four core empirical findings emerge from the numerical results. First, deep cross-border resource integration unlocks massive underutilized rare earth reserves in Outer Mongolia through boosted infrastructure investment efficiency and unified cross-border resource allocation, delivering a permanent downward shift in global rare earth prices and a substantial expansion of China’s domestic high-value rare earth processing capacity centered in Baotou. Second, deep integration yields large net social welfare gains for China, driven by reduced raw material procurement volatility, increased mining and processing sector profits, and revenue from endogenously optimized export taxes on unprocessed mineral exports, while imposing measurable consumer surplus losses on Rest of World manufacturers reliant on cheap Chinese rare earth intermediate products. Third, timely cross-border transportation infrastructure is a non-negotiable prerequisite for capturing integration benefits; a two-year construction delay for Mongolian mining railways severely restricts mineral throughput, eliminates most welfare gains, and creates persistent global rare earth price inflation. Fourth, supply elasticity is a dominant driver of long-run mineral market equilibrium: higher price responsiveness of Outer Mongolian extraction suppresses market price volatility and amplifies China’s supply security advantages under integrated resource governance.

Beyond quantitative results, this analysis reveals dual strategic value of China-Outer Mongolia mineral coordination. Economically, it resolves the mismatch between Outer Mongolia’s abundant untapped critical mineral endowments and China’s world-leading rare earth downstream processing industrial ecosystem, generating mutual industrial synergies for both regions. Strategically, unified cross-border mineral supply drastically lifts China’s supply security index for EV, wind power, and semiconductor raw materials, mitigating external supply chain disruptions from competing global mineral importers. The model’s derived optimal export tax formula further provides a tractable policy tool for regulators to exercise market power over global rare earth trade while balancing domestic industrial input costs.

Several limitations of this analysis create avenues for future research. The model abstracts away geopolitical retaliation risks from RoW nations and ignores dynamic strategic stockpile responses from competing mineral consumers. Subsequent work could incorporate multi-country game-theoretic competition, carbon abatement constraints on Mongolian coal extraction, and heterogeneous downstream manufacturing demand across electric vehicle, wind turbine, and semiconductor sectors. Additionally, extensions may integrate disaggregated green technology adoption curves to capture evolving long-term global demand for rare earth permanent magnets and battery-grade lithium.

In policy terms, this paper delivers clear actionable implications for Sino-Mongolian resource cooperation. Policymakers should prioritize accelerated construction of cross-border railway and energy infrastructure to avoid costly supply bottlenecks, adopt coordinated cross-border mining maintenance regimes to extend infrastructure service life, and deploy flexible export tax instruments calibrated to global demand elasticity to maximize domestic welfare while stabilizing domestic critical mineral input prices. Overall, deep functional economic integration with Outer Mongolia represents a cost-effective, high-impact strategy to secure China’s long-term clean energy material supply chain amid the global low-carbon transition.

\section{Declaration of competing interest}
The authors declare that there are no competing interests related to this research.
\bibliographystyle{elsarticle-num}
\bibliography{ref}
\end{document}